\documentclass[twocolumn:]{pasj02}
\usepackage[switch,mathlines]{lineno} 

\Received{}
\Accepted{}
 
\usepackage{amsmath}

\usepackage[whole]{bxcjkjatype}
\usepackage{url}
\usepackage[T1]{fontenc}
\usepackage{appendix}
\usepackage{textcomp}  
\usepackage{comment}
\usepackage{ulem}

\newcommand{\norv}[1]{{\textcolor{blue}{{\bf Nobu: }{#1}}}}

\newcommand{\norvnew}[1]{{\textcolor{green}{{\bf Nobu: }{#1}}}}

\begin{document} 

\title{ 
XRISM-Subaru views of Abell 754: Energetic ICM Motions Revealed by XRISM/Resolve}


\author{Yuki \textsc{Omiya}\altaffilmark{1} \orcid{0009-0009-9196-4174}%
}
\altaffiltext{1}{Department of Physics, Nagoya University, Aichi 464-8602, Japan}
\email{omiya\_y@u.phys.nagoya-u.ac.jp}
\author{Nobuhiro \textsc{Okabe}\altaffilmark{2,3,4} \orcid{0000-0003-2898-0728}%
}
\altaffiltext{2}{Physics Program, Graduate School of Advanced Science and Engineering, Hiroshima University, 1-3-1 Kagamiyama, Higashi-Hiroshima, Hiroshima 739-8526, Japan}
\altaffiltext{3}{Hiroshima Astrophysical Science Center, Hiroshima University, 1-3-1 Kagamiyama, Higashi-Hiroshima, Hiroshima 739-8526, Japan}
\altaffiltext{4}{Core Research for Energetic Universe, Hiroshima University, 1-3-1, Kagamiyama, Higashi-Hiroshima, Hiroshima 739-8526, Japan}
\author{Kazuhiro \textsc{Nakazawa}\altaffilmark{5,1} \orcid{0000-0003-2930-350X}%
}
\altaffiltext{5}{Kobayashi-Maskawa Institute for the Origin of Particles and the Universe (KMI), Furo-cho, Chikusa-ku, Nagoya, Aichi 464-8601, Japan}
\author{Naomi \textsc{Ota}\altaffilmark{6} \orcid{0000-0002-2784-3652}%
}
\altaffiltext{6}{Department of Physics, Nara Women's University, Nara 630-8506, Japan}
\author{Yuto \textsc{Ichinohe}\altaffilmark{7} \orcid{0000-0002-6102-1441}%
}
\altaffiltext{7}{RIKEN Nishina Center, Saitama 351-0198, Japan}
\author{Shutaro \textsc{Ueda}\altaffilmark{8, 9} \orcid{0000-0001-6252-7922}%
}
\altaffiltext{8}{Faculty of Mathematics and Physics, Institute of Science and Engineering, Kanazawa University, Kakuma, Kanazawa, Ishikawa, 920-1192 Japan}
\altaffiltext{9}{Advanced Research Center for Space Science and Technology (ARC-SAT), Kanazawa University, Kakuma, Kanazawa, Ishikawa, 920-1192, Japan}
\author{Nhan T. \textsc{Nguyen-Dang}\altaffilmark{6} \orcid{0000-0001-6178-7714}%
}



\KeyWords{galaxies: clusters: individual (Abell 754) --  galaxies: clusters: intracluster medium -- X-rays: galaxies: clusters --  turbulence -- large-scale structure of the universe}

\maketitle

\begin{abstract}
We present high-resolution X-ray spectroscopy of the merging cluster Abell~754 using \textit{XRISM}/Resolve. In GO1 phase, \textit{XRISM}/Resolve observed Abell 754 in two deep pointings, targeting the eastern primary core (114~ks) and the middle of the X-ray filamentary structure (190~ks). Spectral fits to full field-of-view data reveal a line-of-sight velocity difference of $656 \pm 35$~km~s$^{-1}$ between the two pointing, corresponding to a bulk Mach number of 0.45$\pm$0.03. Velocity dispersions are measured to be $220^{+26}_{-29}$~km~s$^{-1}$ and $279^{+24}_{-23}$~km~s$^{-1}$ in the eastern and middle pointing, respectively.
Within the eastern core, the velocity dispersion shows spatial variation, reaching $497^{+144}_{-117}$~km~s$^{-1}$ in the southern core with high temperature — among the largest values yet reported in galaxy clusters to date. Narrow-band analysis of the Fe--K complex in this region reveals systematically higher temperatures derived from He-like and H-like Fe line ratio compared to those obtained via broadband fits, indicating multi-phase structures. Two-temperature modeling further separates a cooler core phase from a hotter, shock or turbulence-heated phase whose velocity is blueshifted, similar to that of the middle pointing. These results point to a mixing interface where post-shock gas from the south overlaps, in projection, with cooler core gas, inflating the observed line widths in this region.
Weak-lensing analysis with Subaru/HSC and Suprime-cam confirms that the eastern component is about twice as massive as the western one, consistent with disruption and gas stripping of the latter. The curved morphology of the eastern X-ray core, together with the measured kinematics, is naturally explained by an off-axis, post--core-passage merger that imparts angular momentum and drives large-scale rotational and fallback flows.
\color{black}

\end{abstract}


\section{Introduction}

Galaxy clusters, the most massive gravitationally bound systems in the Universe, assemble hierarchically through mergers and continuous accretion along the cosmic web (e.g., \cite{2005RvMP...77..207V,2011ARA&A..49..409A,2012ARA&A..50..353K}). During cluster mergers, gravitational binding energy of order $10^{64}$~erg is released, a significant fraction of which is transferred into the intracluster medium (ICM) through shocks and turbulence, heating the plasma and generating non-thermal components (e.g., \cite{2007PhR...443....1M,2012MNRAS.421.3375V,2023PASJ...75...37O}). The resulting relativistic particles and amplified magnetic fields produce diffuse synchrotron emission in the form of radio halos and relics (e.g., \cite{2014IJMPD..2330007B,2019SSRv..215...16V}). Constraining how this energy is splitted between thermal and non-thermal components requires direct measurements of gas motions and accurate thermodynamic diagnostics of the hot phase.

High-resolution, non-dispersive X-ray spectroscopy provides a uniquely direct probe of ICM kinematics. 
The pioneering Hitomi/SXS observations of the Perseus core measured relatively low line-of-sight (LOS) velocity dispersions of 150-200~km~s$^{-1}$ and set a benchmark for such studies (\cite{2016Natur.535..117H,2018PASJ...70....9H,2018PASJ...70...10H}). Building on this capability, XRISM/Resolve \citep{2020SPIE11444E..22T,2022SPIE12181E..1SI} now extends precision line diagnostics and velocity measurements to a broader range of environments, including dynamically active, merging clusters. These instruments enable simultaneous constraints on bulk flows, turbulence, and multi-temperature structure using Fe--K line complexes.

Abell~754 (hereafter A754) is a prototypical major merger. Early ROSAT and \textit{Chandra} studies established its highly irregular X-ray morphology and complex temperature structure, including an east--west elongation and strong thermal contrasts indicative of a violent, off-axis encounter (\cite{1995ApJ...443L...9H,2003ApJ...586L..19M}). The eastern side hosts a cold-front–like edge, plausibly reflecting a displaced cool core, whereas the western side appears more disrupted. On larger scales, simulations reproduce many of the observed features with off-axis collisions that impart angular momentum to the gas and can lead to long-lived sloshing and rotation (e.g., \cite{1998ApJ...493...62R,2001ApJ...561..621R,2018ApJS..234....4Z}). At radio wavelengths, A754 hosts a central halo and peripheral relic emission consistent with merger-driven shocks and turbulence (\cite{2001ApJ...559..785K,2004ApJ...605..695G,2009ApJ...699.1883K,2024A&A...690A.222B}). 
Deviations from strict collisional ionization equilibrium have also been discussed in the context of rapid shock heating and subsequent expansion (\cite{2016PASJ...68S..23I}). Weak-lensing maps reveal multiple mass concentrations and gas–dark matter offsets, reinforcing the picture of an on-going merger with significant ram-pressure effects (e.g., \cite{2008PASJ...60..345O}). Individual mass clumps are centered on their respective brightest cluster galaxy (BCG): WISEA~J090832.37$-$093747.3 ($z = 0.054841$, hereafter W-BCG) in the western clump \citep{1991rc3..book.....D}, and WISEA~J090919.21$-$094159.0 ($z = 0.054291$, hereafter E-BCG) in the eastern clump \citep{2009A&A...495..707C}.

\begin{figure}[htb]
 \begin{center}
  \includegraphics[width=8.8cm]{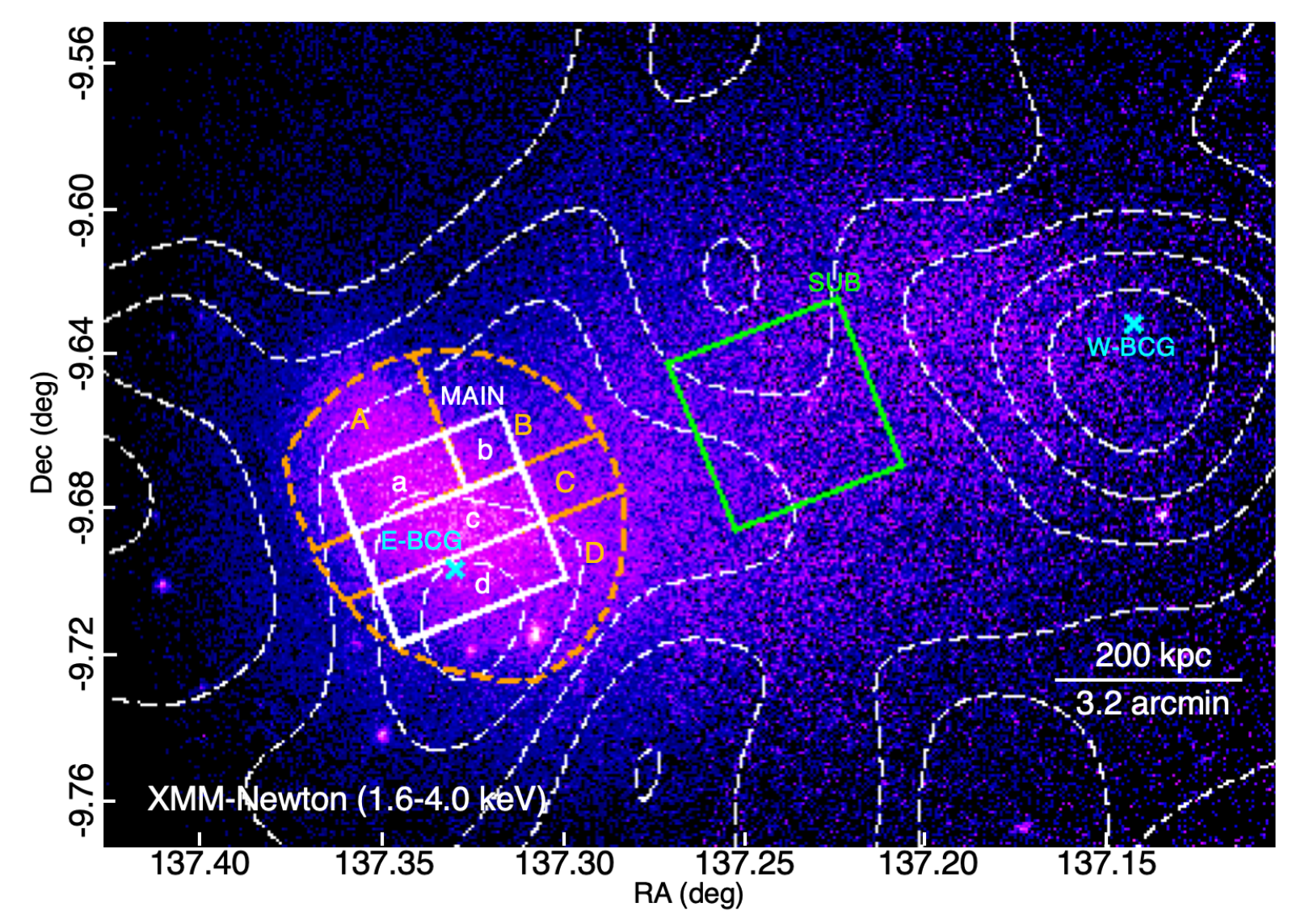}
 \end{center}
\caption{
\textit{XMM-Newton}/EPIC mosaic image of A754 in the 1.6–4.0~keV band (constructed following the procedure described in \cite{2024A&A...689A.173O}), overlaid with color regions used for \textit{XRISM} spatially resolved spectroscopy. The white and green boxes indicate the Resolve FoV for the MAIN and SUB observations, respectively. White rectangles within the white box represent the subgrid areas designated for spectral extraction.
White dashed contours indicate the projected mass distribution from the weak-lensing mass map of \citet{Okabe2025b}, shown at significance levels of 1, 2, 3, 4, and 5$\sigma$.
Cyan crosses mark the positions of the E-BCG ($z_{\rm EB} = 0.054291$) and W-BCG ($z_{\rm WB} = 0.054841$). The dashed orange regions represent the sky regions for which the ICM parameters are derived.
{Alt text: X-ray Image of Abell 754, with right ascension on the horizontal axis and declination on the vertical axis.}}

\label{fig1:image}
\end{figure}

Despite this rich multiwavelength studies, direct spectroscopic measurements of the ICM velocity field in A754 have been lacking. Such measurements are essential to constrain the kinematic decoupling between the collisional gas and the collisionless galaxy/dark-matter components—a hallmark of off-axis merger scenarios—and to overcome the projection and multi-phase biases inherent in broadband temperature maps (e.g., \cite{2004MNRAS.354...10M}).
\color{black}

In this work, we present XRISM/Resolve observations of A754 that deliver spatially resolved measurements of bulk velocities and velocity dispersions across the two dominant X-ray components. We combine broadband and narrow-band (Fe--K) spectral analysis with a spatial–spectral mixing (SSM) treatment of the point-spread function to mitigate cross-contamination between regions. We further leverage complementary wide-field X-ray constraints and weak-lensing information to interpret the dynamics. 
This paper is organized as follows: 
section~2 describes the observations and data reduction; section~3 outlines the spectral analysis strategy; section~4 presents the results from the field-of-view (FoV) and spatially resolved spectroscopy, including line diagnostics and multi-temperature modeling; and section~5 provides a discussion of the ICM kinematics, comparison with galaxies, and the overall merging scenario.
\color{black}

Throughout this paper, we use $H_0=70$ km~s$^{-1}$~Mpc$^{-1}$, $\Omega_{\rm m}=0.3$ flat cosmology, in which $1$ arcmin$=63.1$ kpc at the cluster redshift ($z=0.054$).
All errors are in 1$\sigma$ confidence interval, otherwise noted.

\section{Observations and Data Reduction}

\subsection{XRISM Observations}

In reference to our accompanying paper on weak lensing analysis conducted with Subaru/HSC and Suprime-cam \citep{Okabe2025b}, we have determined using free centroids that A754 represents a major merger with an approximate mass ratio of $2:1$. The main and sub clusters are associated with the eastern and western brightest galaxies (E-BCG and W-BCG), respectively. We thus refer to the eastern and western \textit{XRISM}/Resolve pointings as \texttt{MAIN} and \texttt{SUB} (figure \ref{fig1:image}), respectively. The \texttt{MAIN} observing data (ObsID: 201015010: PI Okabe) centering the cluster gas core (RA: 137.3318, Dec: $-9.68789$; J2000) was obtained on 2024 November 20–22 in the GO1 phase. The \texttt{SUB} pointing data (ObsID: 201016010: PI Okabe) centering the middle of the X-ray filamentary structure (RA: 137.23533, Dec: $-9.65558$) was acquired on 2025 April 22–27 in the GO1 phase. The \texttt{SUB} pointing is selected because the region around the W-BCG has relatively low metallicity, requiring longer exposure for ICM velocity measurements \citep{2016PASJ...68S..23I}, and placing it too near the main core could lead to significant contamination.



\subsection{Data Reduction}
\label{sec:Data Reduction for Resolve}

Data were processed with \texttt{xapipeline} in HEAsoft v6.34 (CALDB v11), following the \textit{XRISM} Quick Start Guide v2.1. 
We applied the standard screening criteria (\texttt{DERIV\_MAX}, \texttt{RISE\_TIME}, and \texttt{STATUS[4]}) and retained only primary high-resolution events (ITYPE\,=\,0), yielding net exposures of 114~ks (\texttt{MAIN}) and 190~ks (\texttt{SUB}). 
Redistribution matrix files (RMFs) were produced with \texttt{rslmkrmf} using the ``L'' setting, which models the electron-loss continuum and escape peaks. 
Ancillary response files (ARFs) were generated with \texttt{xaarfgen} using a mosaic \textit{XMM-Newton}/EPIC image in the 1.6–4.0~keV band, constructed from ObsIDs 0136740101, 0136740201, 0112950401, 0112950301, 0556200101, and 0556200501, as the input surface-brightness map to weight the effective area by the observed morphology. The processing and mosaic construction of this XMM image follow the procedure described in \citet{2024A&A...689A.173O}. The resulting map, used for ARF generation, is shown in figure~\ref{fig1:image}.

We quantify instrumental systematics on the velocity measurements using the post-launch per-pixel $^{55}$Fe calibration data provided for each \textit{Resolve} observation. In these calibrations, the continuously illuminated ``calibration pixel'' and the main-array pixels are irradiated by $^{55}$Fe (Mn K$\alpha,\beta$ near 5.9--6.5~keV) to measure, for every pixel, the energy-scale residual and the line-spread function; the calibration pixel, corrected on the same time grid as the main array, serves as a witness for the intermittent main-array calibration.\footnote{XRISM Resolve energy-scale reports: \url{https://heasarc.gsfc.nasa.gov/FTP/xrism/postlaunch/gainreports}} 
For both A754 pointings (OBSID 201015010 and 201016010), the array-averaged residual energy offsets at 6--7~keV are $<$0.1~eV. 
Adding these in quadrature with the in-orbit energy-scale accuracy of $\pm0.3$~eV (5.4--9.0~keV) yields a conservative $\pm0.32$~eV systematic at 6.4~keV, i.e.\ a line-of-sight bulk-velocity uncertainty of $\sim15$~km~s$^{-1}$ per pointing. 
The measured FWHM across pixels lies in the expected 4.5--4.7~eV range, and the current in-orbit line-spread uncertainty of $\sim$0.15~eV (6--7~keV) maps to only 1--2~km~s$^{-1}$ uncertainty on velocity dispersion of $200$~km~s$^{-1}$. 
Thus, for OBSID~201015010 and 201016010, a systematic uncertainty of bulk motion is $\sim15$~km~s$^{-1}$ and that of velocity dispersion is $<2$~km~s$^{-1}$.
Therefore, in the following results, we quote statistical errors only, without including these systematics.
\color{black}



\begin{figure*}[htpb]
 \begin{center}
  \includegraphics[width=13.0cm]{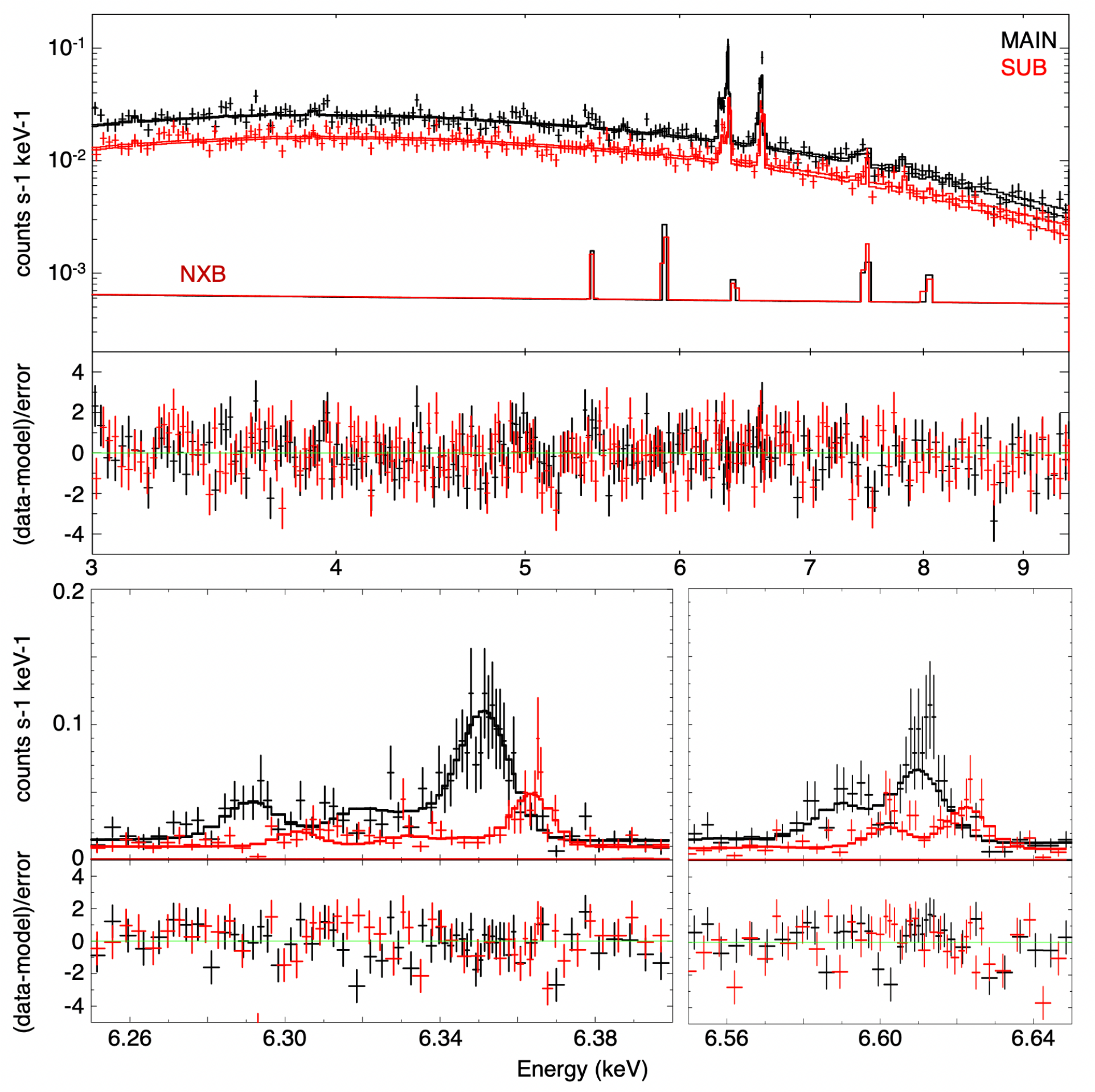} 
 \end{center}
\caption{(Top) Resolve spectra in the 3.0--9.5~keV energy band for MAIN (black) and SUB (red), respectively. The best-fit models are overlaid on each spectrum with NXB models, illustrating the goodness of fit between the observational data and the corresponding model.  Data are binned by a factor of 4~eV for display purposes. (Bottom) Zoom-in of the Resolve spectra around the He-like Fe lines (left) and H-like Fe lines (right).
Data are binned by a factor of 2~eV for display purposes. 
{Alt text: Three line graphs. In the upper panel, the x axis shows the energy from 3.0 to 9.5 kilo electron volt. The y axis shows the count from 0.0002 to 0.2 counts per second and per kilo electron volt, and the residuals of minus 5 to 5 in lower part. In the lower left panel, the x axis shows the energy from 6.25 to 6.40 kilo electron volt. In the lower right panel, the x axis shows the energy from 6.55 to 6.65 kilo electron volt. The y axis shows the count from 0.0 to 0.2 counts per second and per kilo electron volt, and the residuals of minus 5 to 5 in lower part.
\color{black} 
}}
\label{fig2:FoV_spec}
\end{figure*}

\section{Spectral Analysis} 
\label{sec:Spectral Analysis} 

Spectral fitting was performed in \texttt{XSPEC} v12.15.0 \citep{1996ASPC..101...17A} with AtomDB 3.0.9, adopting protosolar abundances from \citet{2009LanB...4B..712L}. The ICM emission was modeled with \texttt{bapec}, assuming collisional ionization equilibrium and including broadening from thermal and velocity dispersion. Free parameters were the temperature, metal abundance, redshift, and (Gaussian) velocity dispersion. Average heliocentric velocity corrections of $+27.0~$km~s$^{-1}$ for \texttt{MAIN} and $-25.0~$kms$^{-1}$ for \texttt{SUB} were applied, as estimated using \texttt{barycen}. Galactic absorption was modeled with \texttt{tbabs} using a fixed column density of $N_{\rm H} = 4.82 \times 10^{20}$~cm$^{-2}$ \citep{2005A&A...440..775K}. 
Non-X-ray background (NXB) spectra were constructed from night-Earth events with \texttt{rslnxbgen}, weighted by the geomagnetic cut-off rigidity, with the Resolve NXB Database (v2). The resulting NXB spectra were then modeled with a power-law and Gaussian lines, and the best-fit model was incorporated as the NXB component in the subsequent spectral fitting.
\color{black}

Given the known low-energy calibration uncertainties of Resolve—leading to systematic temperature offsets relative to \textit{XMM-Newton} and \textit{Chandra} \citep{2025AAS...24541206P}—all fits were restricted to the 3.0--9.5~keV band.

\begin{table}[htpb]
\begin{center}
\tbl{Best-fit Parameters of full array in 3.0--9.5~keV enegy range. }{%
\begin{tabular}{@{}lccc@{}}
\hline\noalign{\vskip3pt}
\multicolumn{1}{c}{Parameter} & Main & Sub  \\ [2pt]
\hline\noalign{\vskip3pt}
 $kT$~(keV) & $8.10^{+0.27}_{-0.26}$ & $9.48^{+0.39}_{-0.38}$  \\
 Abundance~($Z_{\rm{\odot}}$) & $0.44^{+0.03}_{-0.03}$ & $0.29^{+0.03}_{-0.03}$  \\
 Redshift ($\times$10$^{-2}$) & $5.5030^{+0.0086}_{-0.0097}$  & $5.2839^{+0.0085}_{-0.0094}$  \\
  Relative Velocity (km~s$^{-1}$)\footnotemark[$*$] &$+220^{+26}_{-29}$ & $-436^{+26}_{-28}$\\
  Velocity dispersion (km~s$^{-1}$) & $279^{+24}_{-23}$ & $195^{+32}_{-29}$\\
  C-stat/d.o.f.  & $13848/12994$ & $14101/12994$ \\
 [2pt]
\hline\noalign{\vskip3pt}
\end{tabular}
}\label{tab1:Fit_parameter}
\begin{tabnote}
{\hbox to 0pt{\parbox{78mm}{\footnotesize
\par\noindent
\footnotemark[$*$]:  Velocity relative to E-BCG. \\
}\hss}} 
\end{tabnote}
\end{center}
\end{table} 

\section{Results}

\subsection{FoV spectra}
\label{sec:FoV_spectra_of_Resolve}

Figure~\ref{fig2:FoV_spec} shows the Resolve 
full
FoV spectra for the \texttt{MAIN} and \texttt{SUB} pointings, together with the best-fitting single-temperature (1T) \texttt{bapec} models. The corresponding fit parameters are summarized in table~\ref{tab1:Fit_parameter}.

The FoV fittings reveal that the ICM in the \texttt{MAIN} field shows a redshift of  $+220^{+26}_{-29}$~km~s$^{-1}$ relative to the E-BCG, while the \texttt{SUB} field is blueshifted by $-436^{+26}_{-28}$~km~s$^{-1}$.
We note that photon leakage from the \texttt{MAIN} field into the \texttt{SUB} field accounts for less than 0.1\% of the total counts.
The FoV fits also show the LOS velocity dispersions:  
$279^{+24}_{-23}$~km~s$^{-1}$ in \texttt{MAIN} and $195^{+32}_{-29}$~km~s$^{-1}$ in \texttt{SUB}.
\color{black}

In the \texttt{MAIN} spectrum, the 1T model underestimates the intensity of H-like Fe lines, indicating the presence of an additional, hotter plasma component whose enhanced H-like Fe line emission cannot be reproduced by an isothermal description. We examine this possibility in detail in 
sections~\ref{sec:Revisiting_line_diagnostics_with_narrow-band_SSM_fitting} and \ref{sec:2T_modeling}. 
In contrast, the \texttt{SUB} spectrum exhibits a localized deficit near H-like Fe Ly$\alpha_{1}$, whereas the remainder of the Fe--K complex is well reproduced. Similar Fe--K anomalies have been reported in other clusters (e.g., Abell~2029; \cite{2025ApJ...982L...5X}, Coma; \cite{2025ApJ...985L..20X}).

\begin{figure*}[h]
  
  \centering
  \begin{minipage}{0.9\columnwidth}
    \centering
    \includegraphics[width=\columnwidth]{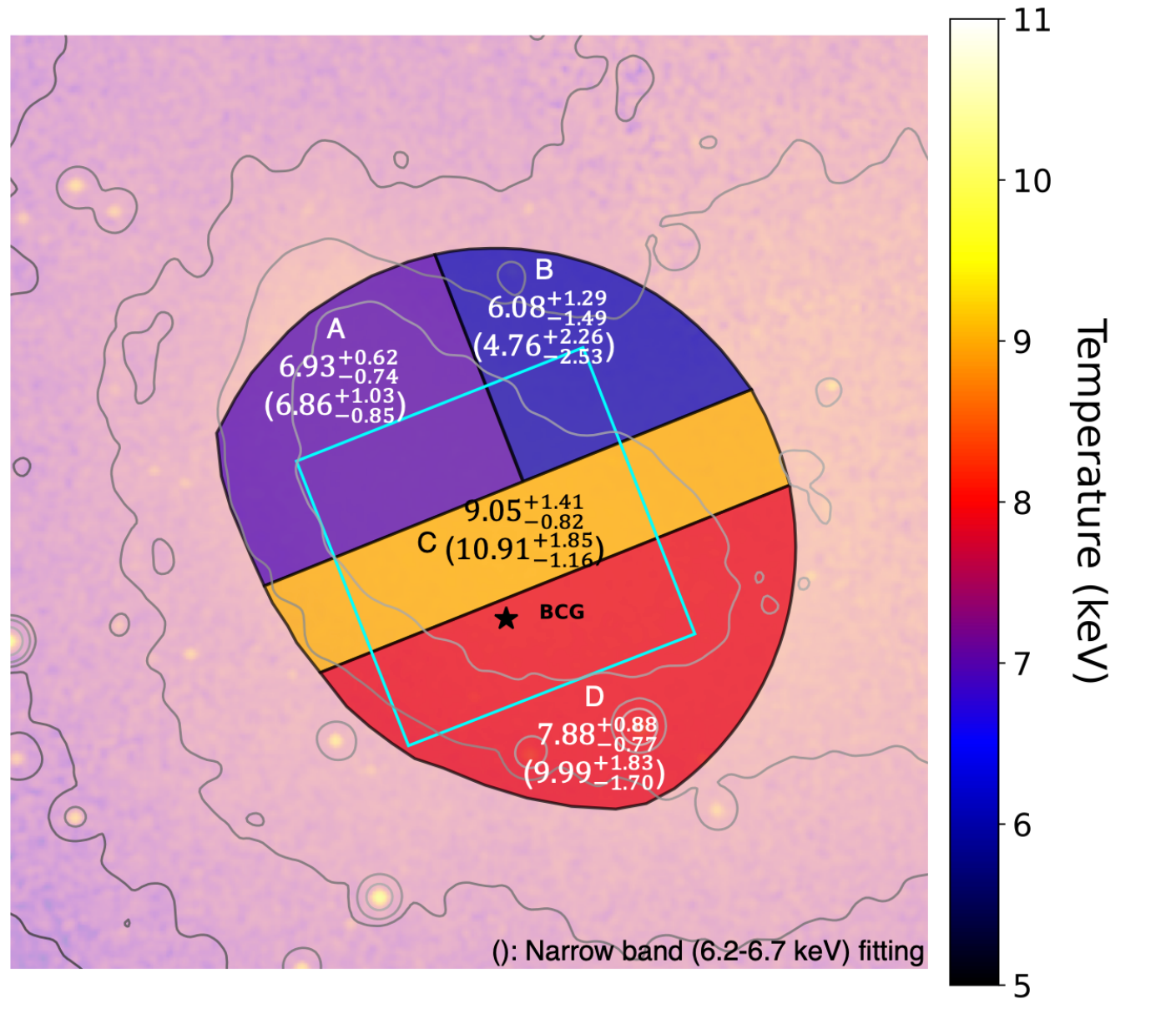}
  \end{minipage}
  \begin{minipage}{0.9\columnwidth}
    \centering
    \includegraphics[width=\columnwidth]{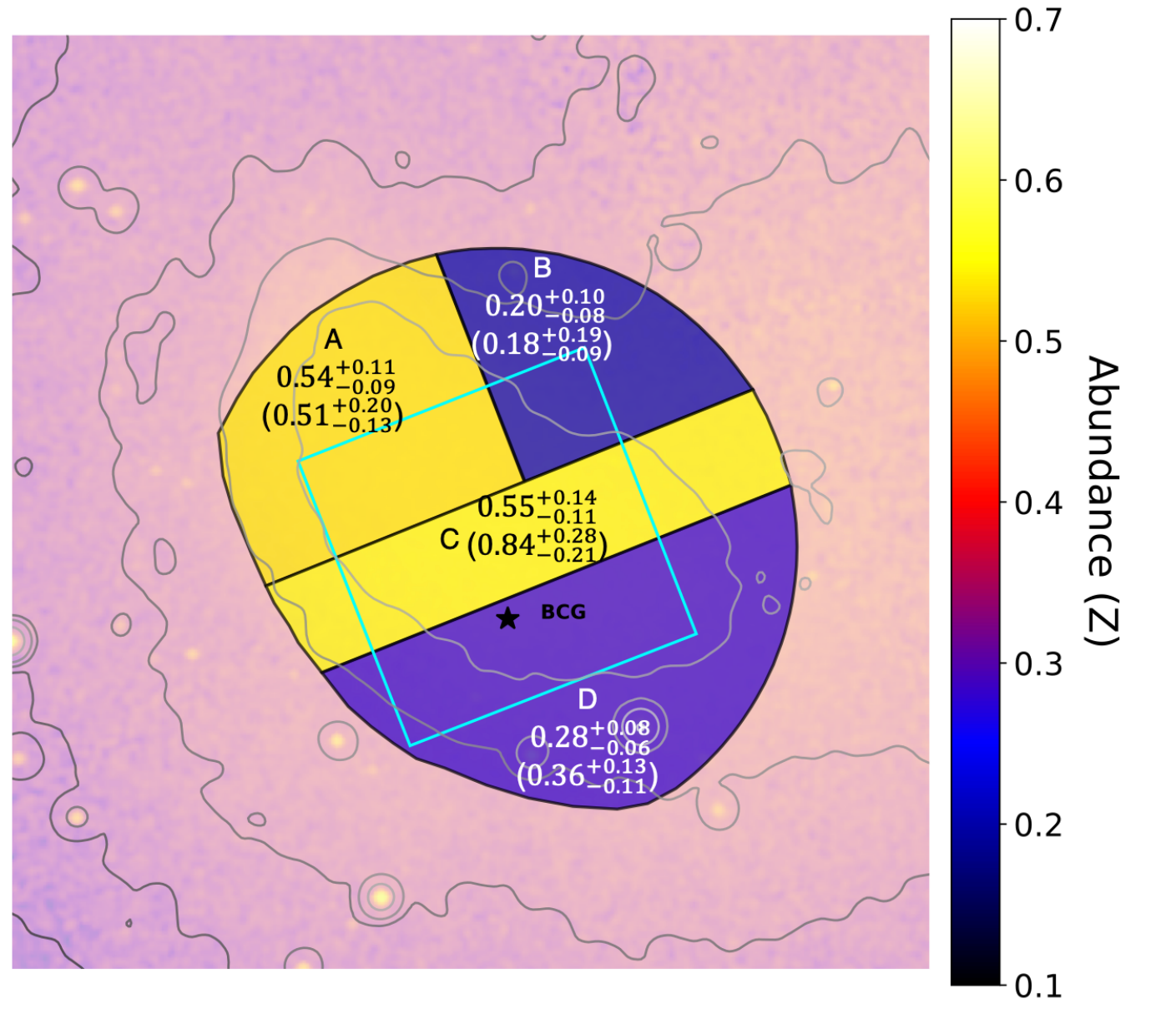}
  \end{minipage}
  
  \centering
  \begin{minipage}{0.9\columnwidth}
    \centering
    \includegraphics[width=\columnwidth]{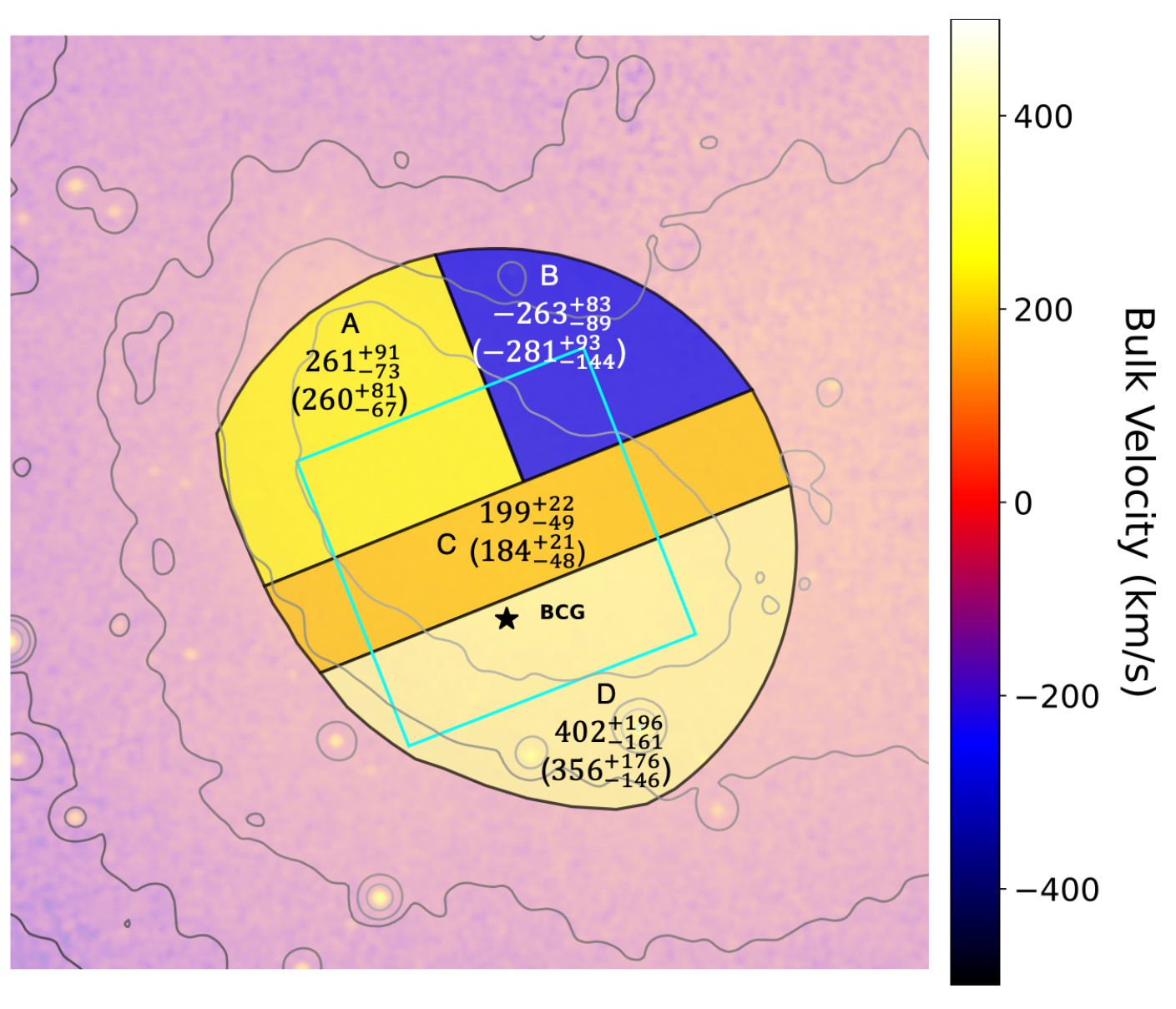}
  \end{minipage}
  \begin{minipage}{0.9\columnwidth}
    \centering
    \includegraphics[width=\columnwidth]{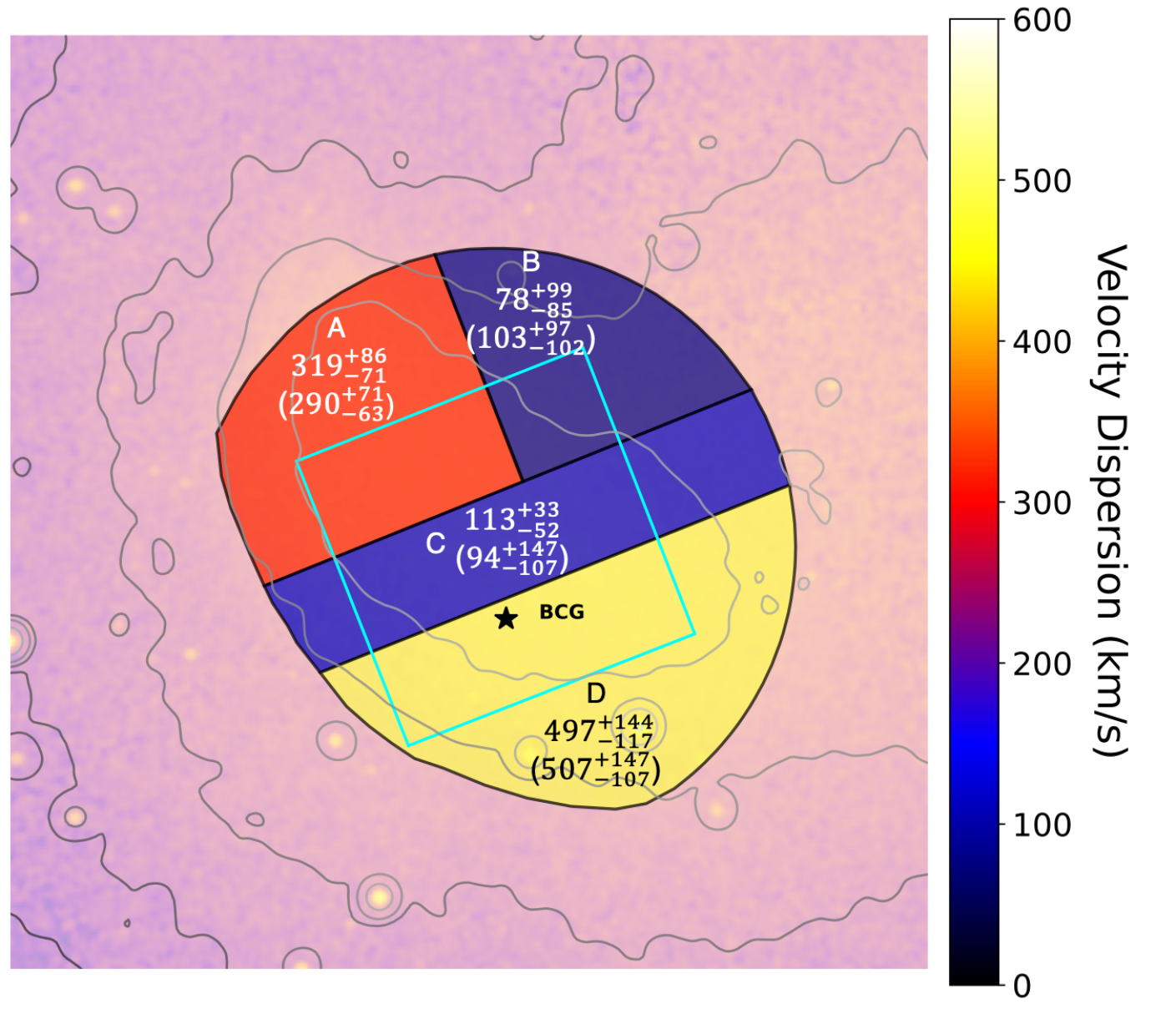}
  \end{minipage}
\begin{flushleft}
\caption{Maps of temperature (top left), abundance (top right), bulk velocity (bottom left), and velocity dispersion (bottom right) for MAIN pointing, based on the spatial regions defined in figure~\ref{fig1:image}. The black star indicates the position of the E-BCG. 
All maps are derived from broad-band fits in the 3.0–9.5~keV range; values in parentheses indicate the corresponding results from the narrow-band fits (6.2–6.7~keV).
Contours represent the X-ray surface brightness, highlighting the correspondence with spatial structures. The cyan box indicates the field of view of the MAIN observation. The units of the color bars are, in order:~keV, solar, km~s$^{-1}$, and km~s$^{-1}$.
{Alt text: Four color maps arranged in two rows and two columns, each with right ascension on the horizontal axis and declination on the vertical axis. The upper left map shows temperature in kilo–electronvolts ranging from 6.0 to 13.0. The upper right map shows abundance in solar units ranging from 0.1 to 0.6. The lower left map shows bulk velocity in kilometers per second ranging from minus 250 to plus 250. The lower right map shows velocity dispersion in kilometers per second ranging from 0 to 500.}}\label{fig3:Resolve_map_ssm}
\end{flushleft}
\end{figure*}

\subsection{Spatially resolved spectroscopy}
\label{sec:Spatially resolved spectroscopy}

To quantify spatial structure within the \texttt{MAIN} pointing while accounting for mixing by the \textit{XRISM} point-spread function (PSF; half-power diameter $\sim$1.3 arcmin), we performed an SSM analysis following the multi-source ARF technique of \citet{2018PASJ...70....9H}. 
We first limit the input sky emission region to within a circular region of radius 3 arcmin (corresponding to the distance of \textit{XRISM} half-power diameter beyond the Resolve field of view) centered on the pointing position, deliberately excluding the very X-ray–faint outskirts.
We divided this area into four sky regions (A--D; see dashed orange lines in figure~\ref{fig1:image}) to isolate physically distinct merger substructures: 
Region~A covers the low-entropy, cool component; 
Region~B is the envelope just outside the cool component; 
Regions~C and D trace the sloshing wake of the cool-core structure, where the gas is expected to be hotter. 
This area was split into two subregions (regions C and D) using the position of the E-BCG as the dividing boundary.
We note that the leak-in contributions from outside the regions~A-D are estimated to be approximately 1\%. 
Since the parameters in the region outside A-D cannot be reliably constrained with the current data, we excluded this region from the spectral modeling.
\color{black}

On the detector plane, the field was divided into regions a–d (see white boxes in figure~\ref{fig1:image}). Regions were chosen to be wider, or similar to, the 
half power diameter
of the PSF so that each spectrum is dominated by its underlying sky region while still sampling gradients across the field. Because each detector region can receive photons originating from any sky region, we generated cross-region ARFs for every (sky, detector) 
regions' pair.
These ARFs were computed using the appropriate attitude solution and were weighted by the 1.6–4.0~keV \textit{XMM-Newton}/EPIC mosaic surface-brightness map (section~\ref{sec:Data Reduction for Resolve}). All spectra from the four detector regions were then fitted simultaneously; parameters associated with the same sky region (temperature, abundance, redshift, velocity dispersion, and normalization of the \texttt{bapec} component) were linked across detector spectra to enforce a single physically consistent solution per sky region.


\begin{figure*}[htb]
 \begin{center}
  \begin{minipage}{0.8\columnwidth}
    \centering
    \includegraphics[width=\columnwidth]{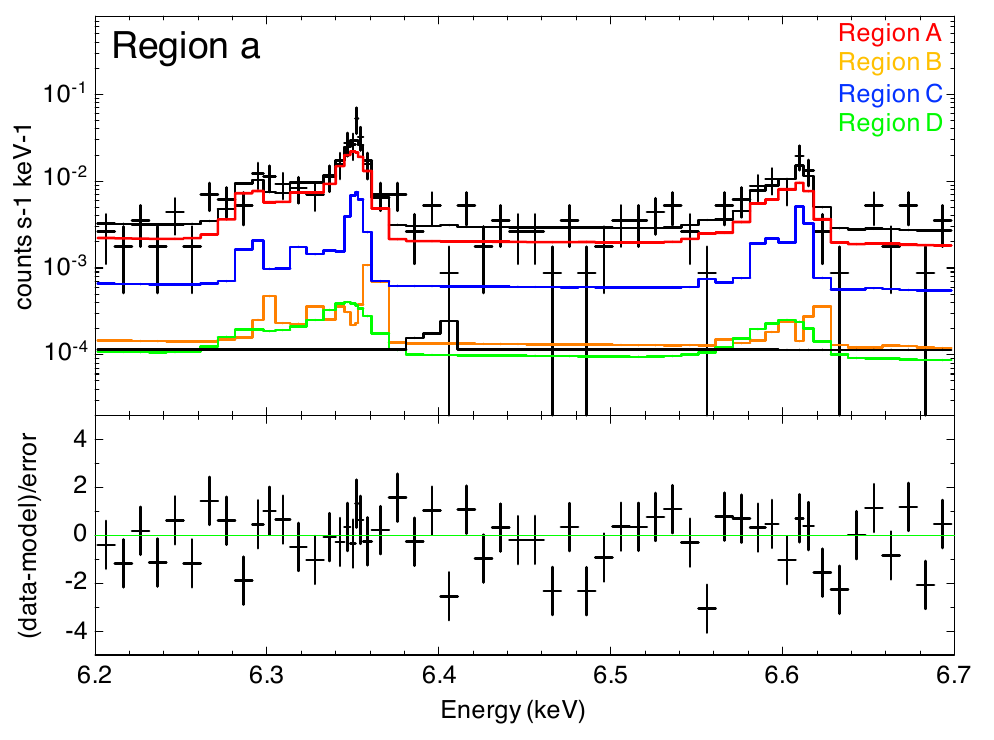}
  \end{minipage}
  \begin{minipage}{0.8\columnwidth}
    \centering
    \includegraphics[width=\columnwidth]{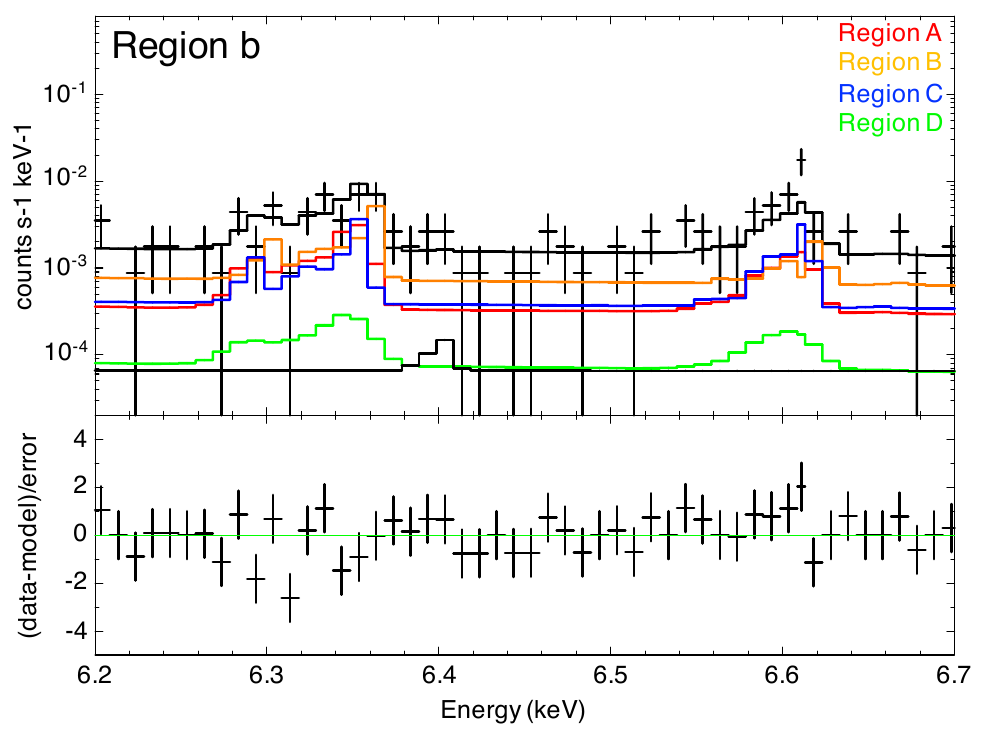}
  \end{minipage}
 \end{center}

 \begin{center}
  \begin{minipage}{0.8\columnwidth}
    \centering
    \includegraphics[width=\columnwidth]{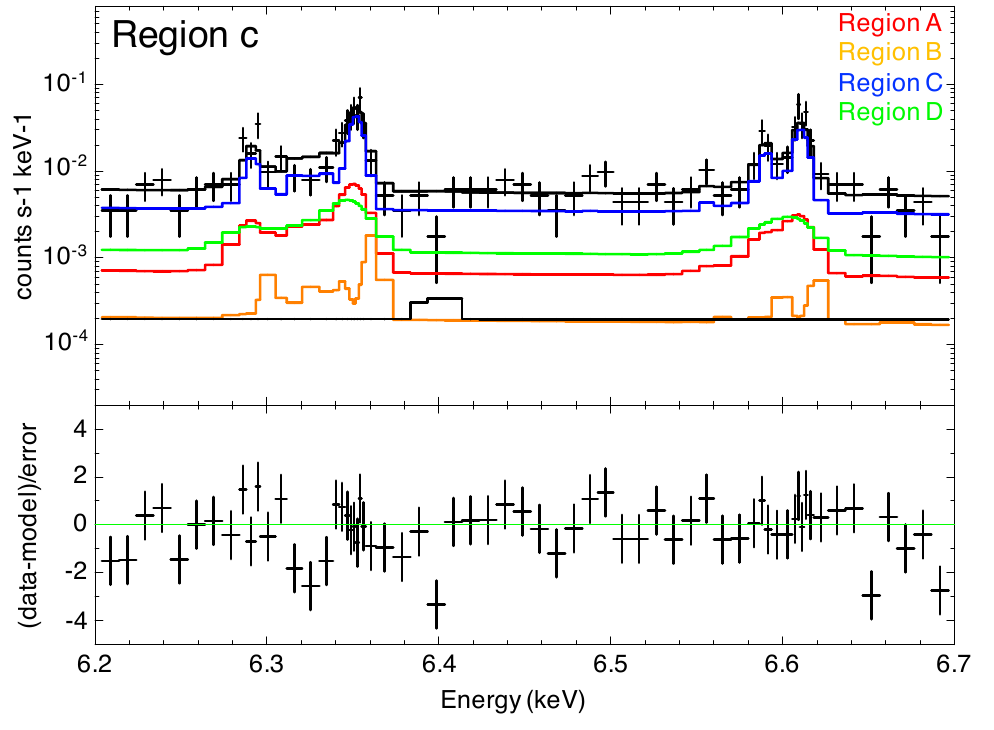}
  \end{minipage}
  \begin{minipage}{0.8\columnwidth}
    \centering
    \includegraphics[width=\columnwidth]{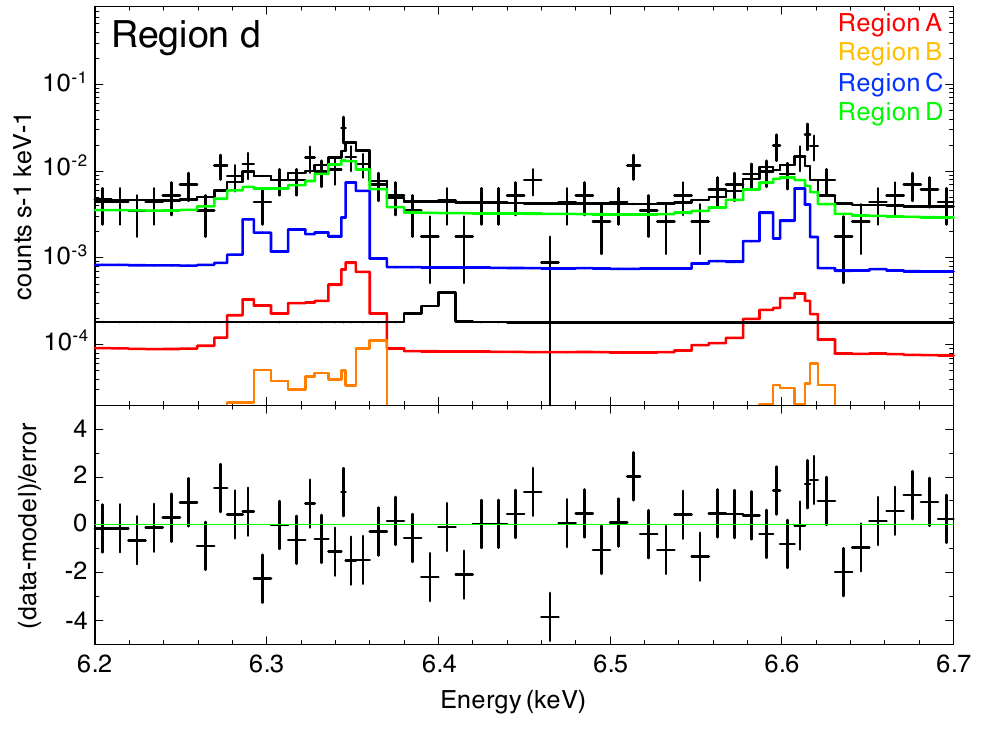}
  \end{minipage}
 \end{center}
\caption{Resolved spectra around the Fe K$\alpha$ lines of detector regions a (upper left), b (upper right), c (lower left), and d (lower right) for the SSM analysis. 
The solid colored lines represent the modeled contributions from individual regions: A (red), B (orange), C (blue), and D (green).
The black lines represent the sum of models.
{Alt text: Two line graphs. In the two panels, the x axis shows the energy from 6.2 to 6.7 kilo electron volt. The y axis shows the count from 0.00002 to 0.8 counts per second and per kilo electron volt, and the residuals of minus 5 to 5 in lower part.}}
\label{fig:spec_regC_D}
\end{figure*}


The resulting maps of temperature, abundance, redshift, and velocity dispersion are shown in figure~\ref{fig3:Resolve_map_ssm}, with corresponding spectra provided in figure~\ref{fig:spec_regC_D}.
From the spatially resolved fits across the \texttt{MAIN} field, region~A is relatively cool at $\sim$7~keV, while region~C is hotter at $\sim$9~keV; both regions have high abundances of $\sim$0.5~solar. In contrast, region~D has a temperature comparable to those in regions~A and C but a lower abundance of $\sim$0.2~solar. These results suggest that regions~A and~C lie near the remnant of the cluster core, with the core center likely located toward the cooler northern side; region~C may have undergone local heating. The ICM–BCG line-of-sight velocity difference is $\sim$+200–300~km~s$^{-1}$ in regions~A and C and $\sim$+400~km~s$^{-1}$ in region~D, potentially tracing coherent bulk flows associated with core sloshing or rotation.
The properties of region~B are discussed later.

The velocity dispersion varies substantially among regions. Region~C is relatively quiescent, with 113$^{+33}_{-52}$~km~s$^{-1}$. Region~A shows intermediate turbulence, with 319$^{+86}_{-71}$~km~s$^{-1}$. Located at the southwestern edge of the \texttt{MAIN} field, region~D, exhibits the most extreme broadening, with 497$^{+144}_{-117}$~km~s$^{-1}$. To our knowledge, this is the largest ICM velocity dispersion reported for a galaxy cluster to date, exceeding previous measurements in other dynamically active systems such as the Coma cluster ($\sim$200~km~s$^{-1}$; \cite{2025ApJ...985L..20X}), Centaurus ($\sim$150~km~s$^{-1}$; \cite{2025Natur.638..365X}), and A2029 ($<170$~km~s$^{-1}$; \cite{2025ApJ...982L...5X}). The spectrum of region~D also shows excess H-like Fe Ly$\alpha$ emission, pointing to an unresolved hotter component—likely shock or turbulence-heated gas—mixing with cooler emission and inflating the measured dispersion, as discussed in section~\ref{sec:2T_modeling}.

One of the noteworthy features is region~B, which is comparatively cool, with $\sim$6~keV, and metal-poor, with an abundance of $\sim$0.2~solar. Although the statistical constraints are limited, its best-fit bulk velocity is $-263^{+83}_{-89}$~km~s$^{-1}$, showing a pronounced blueshift similar to that measured in the \texttt{SUB} field.

In the \texttt{SUB} field, no statistically significant variation in either bulk velocity or velocity dispersion is found across the FoV. 
We then simply devided the region into east and west halves, and fitted 1T model individually (ignoring SSM for simplicity).
The eastern half has a LOS bulk velocity relative to the E-BCG of $-406^{+34}_{-41}$~km~s$^{-1}$ and a velocity dispersion of $183^{+43}_{-38}$~km~s$^{-1}$, while the western half has a bulk velocity of $-402^{+45}_{-43}$~km~s$^{-1}$ and a dispersion of $219^{+49}_{-42}$~km~s$^{-1}$. These values are consistent within their respective uncertainties, indicating no clear spatial variation in the kinematics across the \texttt{SUB} field.

\subsection{Line diagnostics with narrow-band SSM fitting}
\label{sec:Revisiting_line_diagnostics_with_narrow-band_SSM_fitting}
As noted in section~\ref{sec:FoV_spectra_of_Resolve}, the broadband (3--9.5~keV) 1T model fit in \texttt{MAIN} underestimates H-like Fe lines, indicating that a 1T description misses thermal complexity. To emphasize line diagnostics and reduce continuum-driven degeneracies, we performed SSM fitting in a narrow band (6.2--6.7~keV) containing He-like and H-like Fe complex lines, focusing on regions~A--D.
Specifically, the ICM temperature is primarily determined by  intensity ratios of the He-line and H-like Fe lines.

The resultant parameters from the narrow-band fitting are described in parentheses in figure~\ref{fig3:Resolve_map_ssm}. 
The abundance, bulk velocity, and velocity dispersion remain consistent with the broadband fits.
For the temperature, values in regions~C--D are systematically higher than the broadband results, while those in regions~A--B remain consistent within $1\sigma$.
\color{black}
The upward shift in regions~C--D implies an additional hot component needed to reproduce the intensity of the H-like Fe lines. This behaviour is expected in multi-phase plasmas: when several temperature components coexist along the LOS, the broadband spectroscopic temperature is biased toward cooler phases (e.g., \cite{2004MNRAS.354...10M}).
Thus, the combination of a lower broadband temperature and a higher line–ratio temperature is a classic signature of unresolved multi–temperature structure and projection effects. 

We note that uncertainties in PSF mixing can produce apparent discontinuities in the spatial distributions of temperature and abundance, but the derived bulk velocities and velocity dispersions remain essentially unchanged, as seen the results of regions~C and D. Thus, our kinematic results are robust against PSF-related systematics.

\subsection{2T modeling for the southern hot region of the main core}
\label{sec:2T_modeling}

As shown in section~\ref{sec:Spatially resolved spectroscopy}, region~D exhibits a velocity dispersion of $497^{+144}_{-117}$~km~s$^{-1}$, more than $3\sigma$ higher than that of the adjacent region~C. This feature can be naturally explained if gas components with different temperatures and bulk velocities are superimposed along the LOS, causing the composite Fe--K profile to be broader than can be reproduced by a 1T model.

Independent wide-field constraints support this possibility. The \textit{XMM-Newton} temperature map \citep{2024A&A...690A.222B} shows that, while the mean temperature within the \texttt{MAIN} pointing is $\sim$8~keV, a significantly hotter ($\sim$12~keV) sector is present to the south. Region~D lies near the interface between these sectors, suggesting the coexistence or interaction of physically distinct phases.

\begin{figure}[htb]
 \begin{center}
  \includegraphics[width=8.2cm]{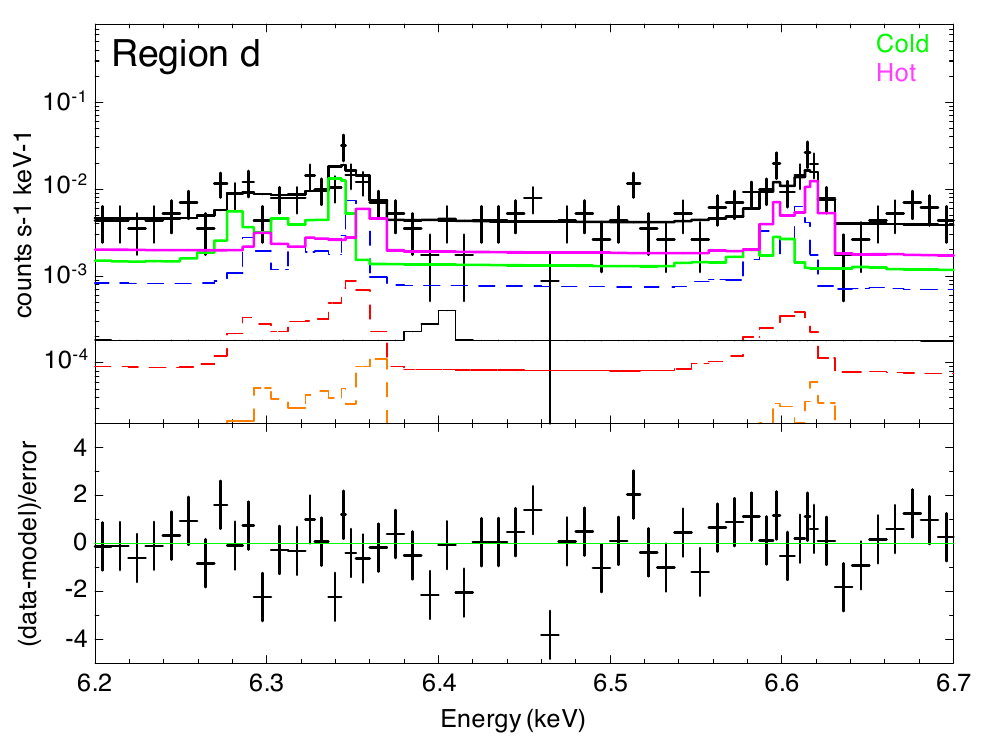}
 \end{center}
\caption{Resolved spectrum around the Fe K$\alpha$ lines of detector region d for the SSM analysis with 2T models. The solid colored lines represent the modeled contributions from individual regions: A (red), B (orange), C (blue), cold component of D (green), and hot component of D (magenta).
The black lines represent the sum of models.
{Alt text: Two line graphs. In the two panels, the x axis shows the energy from 6.2 to 6.7 kilo electron volt. The y axis shows the count from 0.0009 to 0.7 counts per second and per kilo electron volt, and the residuals of minus 5 to 5 in lower part.}}
 \label{fig6:ssm_spec_2T_regD}
\end{figure}

As a simple check, following the method used in \citet{2025arXiv250805067X}, we fitted the He-like Fe lines (6.25–6.40~keV) and H-like Fe lines (6.55–6.65~keV) separately, fixing the temperature at 7.9 keV and the abundance at 0.28~$Z_\odot$.
For region~d, the bulk velocity relative to the E-BCG was measured to be $+476^{+274}_{-240}$~km~s$^{-1}$ (90\% c.l.) from the He-like lines and $-82^{+102}_{-87}$~km~s$^{-1}$ (90\% c.l.) from the H-like lines. These two measurements are inconsistent at more than the $3\sigma$ level, suggesting that cooler and hotter plasma phase along the line of sight may be moving at different velocities.

To further quantify this scenario, we performed a two-temperature (2T) fit to the broad-band (3–9.5~keV) Resolve spectrum of region~d, incorporating SSM and allowing the LOS bulk velocities of the two components to vary independently with respect to the E-BCG (the spectrum is shown in figure~\ref{fig6:ssm_spec_2T_regD}). The resulting temperatures and abundances are $kT_{\rm cool}\simeq 4.6^{+1.4}_{-1.1}$~keV with $Z_{\rm cool}\!\sim\!0.2~Z_\odot$ for the cool phase, and $kT_{\rm hot}\simeq 16.2^{+3.1}_{-2.5}$~keV with $Z_{\rm hot}\!\sim\!0.6~Z_\odot$ for the hot phase. The emission-measure ratio is $\mathrm{EM}_{\rm hot}/\mathrm{EM}_{\rm cool} \approx 0.6$. In velocity space, the cool phase is redshifted by $+697^{+69}_{-98}$~km~s$^{-1}$ and the hot phase is blueshifted by $-170^{+72}_{-66}$~km~s$^{-1}$ relative to the E-BCG, yielding a phase-to-phase velocity separation of $867^{+95}_{-122}$~km~s$^{-1}$. The velocity of the hot component shows a similar blueshift to that measured in the \texttt{SUB} region, and explicitly modeling the hot phase removes the systematic positive residuals on the high-energy side of the Fe--K complex. Therefore, the extreme line width inferred from a 1T fit is more likely an artifact caused primarily by LOS superposition, rather than evidence for unusually strong turbulence.


\section{Discussion}

\subsection{Turbulent and bulk pressure fractions}
\label{sec:bulk_turb_kinetic_pressure}

The FoV-integrated fits (section~\ref{sec:FoV_spectra_of_Resolve}) measure a LOS bulk-velocity difference of $656\pm35$~km~s$^{-1}$ between \texttt{MAIN} and \texttt{SUB}. 
For \texttt{MAIN}, $kT=8.10$~keV implies $c_s=1456$~km~s$^{-1}$ (with $\gamma=5/3$ and $\mu=0.60$), giving a subsonic bulk Mach number of $\mathcal{M}_{\rm bulk}=0.45\pm0.04$. 
This result indicates coherent, large-scale bulk motions across the merger interface.

In \texttt{MAIN}, the LOS velocity dispersion is $279^{+24}_{-23}$~km~s$^{-1}$; in \texttt{SUB}, $195^{+32}_{-29}$~km~s$^{-1}$. 
Assuming isotropy, we convert the LOS dispersion to the 3D turbulent Mach number via $\mathcal{M}_{\rm 3D}=\sqrt{3}\,\sigma_z/c_s$ and estimate the nonthermal pressure (NT) fraction ($P_{\rm NT}/P_{\rm tot}$ ) due solely to turbulence \citep{2019A&A...621A..40E} as
\begin{equation}
\frac{P_{\rm NT}}{P_{\rm tot}}=\frac{\mathcal{M}_{\rm 3D}^2}{\mathcal{M}_{\rm 3D}^2+3/\gamma}\,,
\end{equation}
yielding $P_{\rm NT}/P_{\rm tot} = 5.7^{+0.9}_{-0.8}\%$ in \texttt{MAIN} and $2.4^{+0.8}_{-0.7}\%$ in \texttt{SUB} (table~\ref{table:f_NT}). 
Across \texttt{MAIN}, the turbulence level is heterogeneous: Region~C is relatively quiescent ($0.9^{+0.8}_{-0.5}\%$), Region~A is intermediate ($8.4^{+4.1}_{-3.5}\%$), and region~D is very high ($16.4^{+7.5}_{-7.6}\%$). 
We note, however, that in region~D a 2T description (section~\ref{sec:2T_modeling}) removes the high-energy residuals in Fe--K and reduces the need for extreme intrinsic broadening, indicating that the 1T-based dispersion should be regarded as an upper bound where multi-temperature and LOS superposition are important.


The above $P_{\rm NT}$ refers to the turbulent component only. 
Coherent bulk flows do not act as pressure locally, but their ram pressure and kinetic energy density are dynamically relevant. We define an instructive, density-independent ratio is the ram-to-thermal pressure ($P_{\rm ram}/P_{\rm th}$):
\begin{equation}
\frac{P_{\rm ram}}{P_{\rm th}}=\frac{\frac{1}{2}\rho v_{\rm bulk}^2}{n kT}=\frac{\mu m_p v_{\rm bulk}^2}{2kT}=\frac{\gamma}{2}\,\mathcal{M}_{\rm bulk}^2.
\end{equation}
Here, $\rho$ is the gas mass density, $n$ is the number density, 
$m_p$ is the proton mass, and $v_{\rm bulk}$ is the bulk velocity difference. 
We first estimate the ram-to-thermal pressure ratio for the merger-scale bulk flow. Adopting the line-of-sight velocity difference between \texttt{SUB} and \texttt{MAIN} from the full FoV fitting, we obtain $16.7^{+2.1}_{-2.0}\%$ (table~\ref{table:f_NT}). Because the X-ray morphology exhibits a filamentary structure, plane-of-sky components of the motion are expected; hence, this value should be regarded as a lower limit. Although the turbulent support is modest (a few percent), this merger-driven bulk motion already carries a substantial kinetic load that can promote mixing and transport across the interface. We further evaluate the relative velocity between the western subcluster gas and its BCG (\texttt{SUB}–W-BCG), yielding $2.4^{+0.5}_{-0.5}\%$. This smaller, yet non-zero ratio indicates that the western subcluster gas is moving slightly differently from its BCG, implying a partial decoupling between the collisional ICM and the collisionless galaxy/dark-matter component. Such a discrepancy is consistent with the disrupted and stripped nature of the western core.

In contrast, the eastern core exhibits sloshing-induced bulk motions, which we probe through the velocity difference between Region~D and the E–BCG (\texttt{D}–BCG). This comparison effectively measures the ram pressure of the sloshing gas relative to the central galaxy. The corresponding ram-to-thermal ratio, $8.4^{+7.0}_{-4.4}\%$, while smaller than that for the SUB–MAIN case, demonstrates that sloshing motions can also provide a dynamically relevant contribution, sustaining mixing and redistributing entropy within the core.

\begin{table}[]
    \centering
    \begin{tabular}{c|c}
    Region & $P_{\rm NT}/P_{\rm tot}$\\
    \hline
    \texttt{MAIN} & $5.7^{+0.9}_{-0.9}\%$ \\
    \texttt{SUB}  & $2.4^{+0.8}_{-0.7}\%$ \\
     A    & $8.4^{+4.1}_{-3.5}\%$\\
     B    & $0.6^{+0.6}_{-0.2}\%$\\
     C    & $0.9^{+0.8}_{-0.5}\%$ \\
     D (1T)  & $16.4^{+7.5}_{-7.6}\%$ \\
     \hline
     \hline
     Region & $P_{\rm ram}/P_{\rm tot}$\\
     \hline 
     \texttt{SUB}-\texttt{MAIN}  & $16.7^{+2.1}_{-2.0}\%$ \\
       \texttt{SUB}-W-BCG   & $2.4^{+0.5}_{-0.5}\%$\\
     \texttt{D (1T)}-\texttt{E-BCG}  & $8.4^{+7.0}_{-4.4}\%$\\
     \hline
    \end{tabular}
    \caption{Summary of the nonthermal ( $P_{\rm NT}/P_{\rm tot}$) and ram pressure ( $P_{\rm ram}/P_{\rm tot}$) fractions.}
    \label{table:f_NT}
\end{table}

\color{black}

\subsection{Merging scenario}
\label{sec:Merging_scenario}
With the E-BCG as the reference frame, our spatially resolved spectroscopy shows that the ICM in \texttt{MAIN} is redshifted by $\sim$220~km~s$^{-1}$, whereas the ICM in \texttt{SUB} is blueshifted by $\sim$440~km~s$^{-1}$, yielding a LOS separation of $\sim$660~km~s$^{-1}$. 
In contrast, the velocity difference between the E-BCG and the W-BCG is only $156$~km~s$^{-1}$.
This disparity points to a kinematic decoupling between galaxies and gas: galaxies (and their host dark matter) retain their orbital motion, while the collisional ICM is affected by ram pressure during the encounter and can show different redshift values (e.g., \cite{2004ApJ...606..819M}).
As discussed in Section 3.3 of \citet{Okabe2025b}, the spectroscopic redshift distribution of member galaxies demonstrates that the E-BCG and W-BCG reliably trace the systemic velocities of the eastern main cluster and the western subcluster, respectively.  
We therefore adopt the BCG redshifts as representative of the dynamical states of the two mass components.
Assuming that the position and velocity of the eastern X-ray core coincide with those of the E-BCG at the initial state, the bulk motion of the eastern gas core is likely to be triggered by the hydrodynamical interaction at core passage. The redshifted bulk velocity was induced by a subcluster that passed closer to us along the line of sight than the cluster center.

\begin{figure}[htb]
 \begin{center}
  \includegraphics[width=7cm]{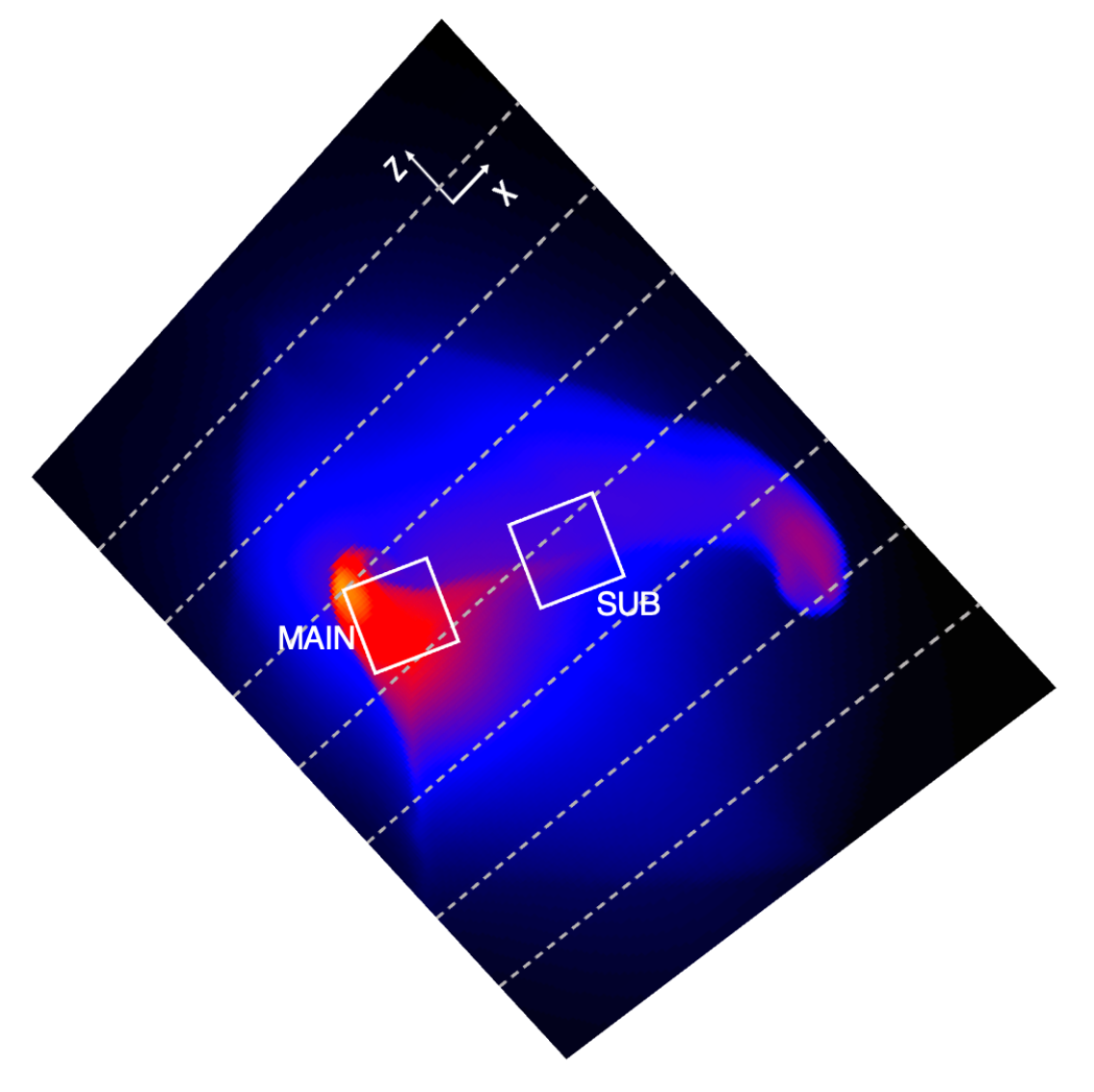} 
 \end{center}
\caption{Projected surface brightness distribution of an example case provided by the Galaxy Cluster Merger Catalog \citep{2016arXiv160904121Z,2018ApJS..234....4Z}. Time epoch after 220~Myr of core crossing initiation are shown. The initial parameter of the mass ratio is set to 3:1. The initial impact parameter is set at 500~kpc in the z-axis direction, and the two clusters approach each other by moving along the x-axis. For display, the simulation is rotated so that the merger plane is inclined with respect to the sky, corresponding to the viewing geometry relevant for A754. The white dashed lines mark parallels to the $x$–$z$ plane.
{Alt text: Projected brightness map of a simulated cluster merger, with white dashed lines drawn parallel to the x–z plane for visual reference.}
}
\label{fig5:Zuhone_simulation}
\end{figure}

Weak-lensing analysis constraints from Subaru/HSC \citep{Okabe2025b} indicate that the eastern mass component centered on the E-BCG is roughly twice as massive as the western counterpart near the W-BCG. The absence of a distinct X-ray core in the western subcluster suggests that its gaseous core has been largely disrupted or stripped. Although this is primarily a morphological inference, the marked flattening of the entropy profile toward the center in the western subcluster is also consistent with this picture \citep{2024A&A...690A.222B}.
Such displacement of the core from the potential minimum may not only manifest as sloshing, but can also evolve into rotational flows once angular momentum is imparted during the offset collision, as discussed below (section~\ref{sec:simulation_comparison}).

    \label{fig:placeholder}

The eastern X-ray core retains a pronounced curved, Tadpole-like 
morphology: the coldest gas lies at the northern tip and the X-ray peak lies at the southern region, and  they are offset from the BCG position \citep{2003ApJ...586L..19M}. Such an offset indicates that the gaseous core has been displaced from the potential minimum and is now sloshing—i.e., undergoing bulk oscillatory motion—within the cluster potential. The eastern core also shows markedly lower entropy than its surroundings, which more naturally reflects the outward transport of pre-existing low-entropy gas, rather than due to adiabatic expansion \citep{2004ApJ...615..181H}. Its asymmetric extension along the merger axis is consistent with partial mixing between low-entropy gas originating in the east and gas stripped from the western subcluster; the latter is likely to expand and cool as it is transported outward, further shaping the observed temperature and entropy structure. 
Taken together, these signatures are hallmarks of off-axis mergers that impart angular momentum and trigger long-lived gas sloshing, and may ultimately evolve into rotational flows, as discussed in section~\ref{sec:simulation_comparison} (e.g., \cite{2006ApJ...650..102A,2013ApJ...764...60R}).

\subsection{Simulation comparison}
\label{sec:simulation_comparison}

To interpret the observed structure, we compare with the Galaxy Cluster Merger Catalog simulations \citep{2018ApJS..234....4Z}. Figure~\ref{fig5:Zuhone_simulation} shows a $3{:}1$ mass-ratio merger with an initial impact parameter of 500~kpc (initial separation is 2.6 Mpc) at $220$~Myr after core passage. The simulation snapshot is selected based on a joint analysis of spectroscopic redshifts of cluster galaxies, weak-lensing masses, and X-ray information can constrain the merger trajectory \citep{Okabe2025b}. They found that the merger mass ratio is $\sim 2:1$, the impact parameter, $b$, is $\sim0.77$ Mpc at the initial separation of 2 Mpc, the inclination angle of the merger plane from the line-of-sight is only $20\pm19$ degrees.

The primary distinction between the numerical simulation and the observed cluster lies in the gas core of the subcluster: it has already been obliterated in observation, whereas it remains intact in the simulation. This could stem from variations in the size of the cool core or the impact parameter at the outset, but it would not be substantially important in our discussion.
At this epoch the cores have undergone an offset collision, imparting angular momentum to the ICM and inducing rotational flow; the resultant Tadpole-like primary core closely resembles the \texttt{MAIN} region in A754. 
The observed velocity structure, with the subcluster motion approaching turnaround, is consistent with this dynamical phase.
In the simulation, both BCGs are near the apocenters of their post-passage orbits, producing a small LOS velocity offset between them (consistent with the observed 156~km~s$^{-1}$), while the gas---having acquired angular momentum---follows diverging trajectories. 
When the merger plane is rotated to an intermediate inclination relative to the sky—corresponding to the viewing geometry adopted in figure~\ref{fig5:Zuhone_simulation}—these motions project into a large LOS separation, naturally explaining the $\sim$600~km~s$^{-1}$ difference between \texttt{MAIN} and \texttt{SUB}.
The \texttt{SUB} region lies in the wake of the western subcluster, where post-collision gas streams are deflected and decelerated. 
The western subcluster core itself is expected to carry a larger line-of-sight velocity relative to the eastern main cluster than that observed at the \texttt{SUB} position.
Since the gas streamlines are deflected along the shape of the body, the velocity vectors of the gas particles gradually change in the direction of the body surface (e.g., \cite{landau1987fluid,2007ApJ...663..816A}). Such a component would partially contribute to the observed line-broadening. 
 

The sharp contrast between the modest dispersion in region~C and the extreme broadening in region~D can be explained by numerical simulations of off-axis mergers (e.g., \cite{2011ApJ...743...16Z,2016ApJ...821....6Z,2006ApJ...650..102A}), which show that rotational and fallback flows around the primary core generate oppositely directed velocity components. When projected along certain sightlines, these components overlap, producing apparent dispersions much larger than the intrinsic turbulence.

In addition, our spectroscopy reveals evidence for a mixing interface: post-shock or turbulence-heated (e.g., \cite{Fujita_2004}), high-temperature gas from the southern sector appears to overlap, in projection, with cooler core plasma redshift. The hot phase in region~D shows a blueshift consistent with that of \texttt{SUB}, suggesting that they may share the same kinematic system. However, the spatial continuity of this hot component and the detailed velocity gradient remain unconstrained due to the gap between the current pointings, motivating future \textit{XRISM} observations to bridge this region.

\subsection{Assessment of NEI state}
We also investigated departures from collisional ionization equilibrium (CIE) in light of the inferred gas dynamics. X-ray mapping of A754 \citep{2004ApJ...615..181H} suggests that the eastern cluster core remnant contains a concentration of cool gas composed of unshocked main-cluster gas mixed with low-entropy gas stripped from the western subcluster during core passage. This stripped gas is deflected by the main core and driven outward, where it expands into lower-pressure regions and cools adiabatically, producing the largest accumulation of low-entropy material at the tip of the core remnant. Such outward transport and adiabatic cooling provide conditions in which the electron temperature may decrease faster than the recombination timescale, potentially leading to a transient non-equilibrium ionization (NEI) state in which the Fe charge-state distribution reflects an ionization temperature exceeding the electron temperature \citep{1984Ap&SS..98..367M,2008PASJ...60L..19A,2010PASJ...62..335A}. At $\sim$220 Myr after core passage (figure \ref{fig5:Zuhone_simulation}), however, a representative post-shock density $n_{\rm e}=10^{-3}$ cm$^{-3}$ yields $n_{\rm e}t \simeq 6.7\times10^{12}$ cm$^{-3}$ s, near or above the CIE threshold in the Fe-K regime, suggesting that NEI effects cannot be the main reason for the observed line ratio difference.

\section{Conclusions}

Our \textit{XRISM}/Resolve observations of A754 have provided a comprehensive view of the gas kinematics and thermodynamics in this archetypal  off-axis merger. With two deep pointings on the eastern primary core and the middle of the X-ray filamentary structure, we directly measured a line-of-sight velocity difference of $656 \pm 35$ km s$^{-1}$, corresponding to a bulk Mach number of $\mathcal{M}_{\rm bulk}\simeq0.45$. This implies a ram-to-thermal pressure ratio of $\sim$0.3, demonstrating that subsonic but coherent bulk flows carry a dynamically significant fraction of kinetic energy. The measured turbulent velocity dispersions are $279^{+24}_{-23}$ km s$^{-1}$ in the eastern core  (or \texttt{MAIN}) and $195^{+32}_{-29}$ km s$^{-1}$ in the middle of the X-ray filamentary structure (or \texttt{SUB}), corresponding to non-thermal pressure fractions of a few percent, comparable to or smaller than values reported in other clusters. 

Within the eastern core, however, the velocity dispersion is far from uniform. Region-to-region variations are pronounced, with the southern sector reaching $497^{+144}_{-117}$ km s$^{-1}$ --- the largest ICM velocity dispersion reported in clusters to date. Such an extreme value cannot be explained by turbulence alone. Narrow-band analysis of the Fe–K complex reveals systematically higher line-ratio temperatures than broadband fits, consistent with the presence of multi-phase structure. Two-temperature modeling explicitly separates a cooler $\sim$5 keV phase from a hotter, shock- or turbulence-heated $\sim$16 keV phase. The hot component shows a blueshift similar to that of the middle of the X-ray filamentary structure, indicating that the large apparent broadening arises from superposition at a mixing interface where post-shock gas from the south overlaps, in projection, with cooler core plasma. This finding highlights the importance of disentangling multi-phase structure and projection effects when interpreting velocity 
dispersions.

Weak-lensing analysis with Subaru/HSC and Suprime-cam confirms that the eastern mass component is about twice as massive as the western one, consistent with the disruption and gas stripping of the latter. The curved morphology of the eastern X-ray core, combined with its measured kinematics, is naturally explained by an off-axis, post–core-passage merger that injected angular momentum into the system. In this scenario, rotational and fallback flows around the primary core generate oppositely directed velocity components, which, when projected along certain sightlines, produce apparent line broadening far in excess of the intrinsic turbulence. 
Taken together, these results establish a coherent dynamical picture of A754 as a post–core-passage system viewed at moderate inclination, highlighting how merger-driven gas dynamics redistribute and thermalize gravitational energy in galaxy clusters.

\begin{ack}
Y.O. would like to take this opportunity to thank the ``Nagoya University Interdisciplinary Frontier Fellowship'' supported by Nagoya University and JST, the establishment of university fellowships towards the creation of science technology innovation, Grant Number JPMJFS2120.
Y.O. was supported by the Sasakawa Scientific Research Grant from The Japan Science Society.
This work was supported by JSPS KAKENHI grant numbers 
JP25K07368 (N.O.), JP20H00157 (K.N.), and JP25K23398 (S.U.).
N.O. acknowledges partial support by the Organization for the Promotion of Gender Equality at Nara Women's University.
S.U. acknowledges support by Program for Forming Japan's Peak Research Universities (J-PEAKS) Grant Number JPJS00420230006. 
This work made use of data from the Galaxy Cluster Merger Catalog (http://gcmc.hub.yt).
\end{ack}







\bibliographystyle{pasj} 
\bibliography{pasj} 

\end{document}